\documentclass{cta-author}

{}
{}
{}

\usepackage{graphics} 
\usepackage{epstopdf}
\usepackage{subfigure}
\usepackage{setspace}
\usepackage{adjustbox}

\begin{document}

\title{Exploiting Complexity in Pen- and Touch-Based Signature Biometrics}

\author{\au{Ruben Tolosana$^{1\corr}$}, \au{Ruben Vera-Rodriguez$^{1}$}, \au{Richard Guest$^{2}$}, \au{Julian Fierrez$^{1}$} and \au{Javier Ortega-Garcia$^{1}$}}

\address{\add{1}{Biometrics and Data Pattern Analytics Lab, Universidad Autonoma de Madrid, 28049 Madrid, Spain}
\add{2}{School of Engineering and Digital Arts, University of Kent, UK}
\email{ruben.tolosana@uam.es}}

\begin{abstract}
Biometric signature verification has been traditionally performed in pen-based office-like scenarios using devices specifically designed for acquiring handwriting. However, the high deployment of devices such as smartphones and tablets has given rise to new and thriving scenarios for signature biometrics where handwriting can be performed using not only a pen stylus but also the finger via touch interaction. Some preliminary studies have highlighted the challenge of this new scenario and the necessity of further research on the topic. The main contribution of this work is to propose a new on-line signature verification architecture adapted to the signature complexity in order to tackle this new and challenging scenario. Additionally, an exhaustive comparative analysis of both pen- and touch-based scenarios using our proposed methodology is carried out along with a review of the most relevant and recent studies in the field. Significant improvements of biometric verification performance and practical insights are extracted for the application of signature verification in real scenarios.
\end{abstract}

\maketitle

\section{Introduction}\label{sec1}

On-line signature verification has been studied in depth in recent years proving to be one of the most reliable and convenient biometric systems in many relevant sectors such as security, e-government, healthcare, education, banking, and insurance regardless of the age of the user~\cite{moises_ACM,Guest20061098}. In~\cite{2015_ICCST_SignatureRobTemp_RubenT}, an approach for irreversible signature template generation was proposed, avoiding the use of sensitive information related to \textit{X} and \textit{Y} coordinates and their derivatives on the biometric system. As a consequence, more robust signature verification systems were developed against cyberattacks as critical information was not stored anywhere. That approach achieved results below 7.0\% and 1.0\% Equal Error Rate (EER) using the pen as the writing input for skilled and random forgeries, respectively. Other challenges of practical importance such as the template aging (i.e., the gradual decrease in a system performance due to the user changes across time) and the input device interoperability were recently studied in~\cite{galbally13PONEagingSignature,2019_IETB_Aging_Tolosana,sae2014online,iet:/content/journals/10.1049/iet-bmt.2013.0044,2015_IEEEAccess_InterSign_Tolosana}.

\begin{figure*}[tb]
  \centering
    \includegraphics[width=\linewidth]{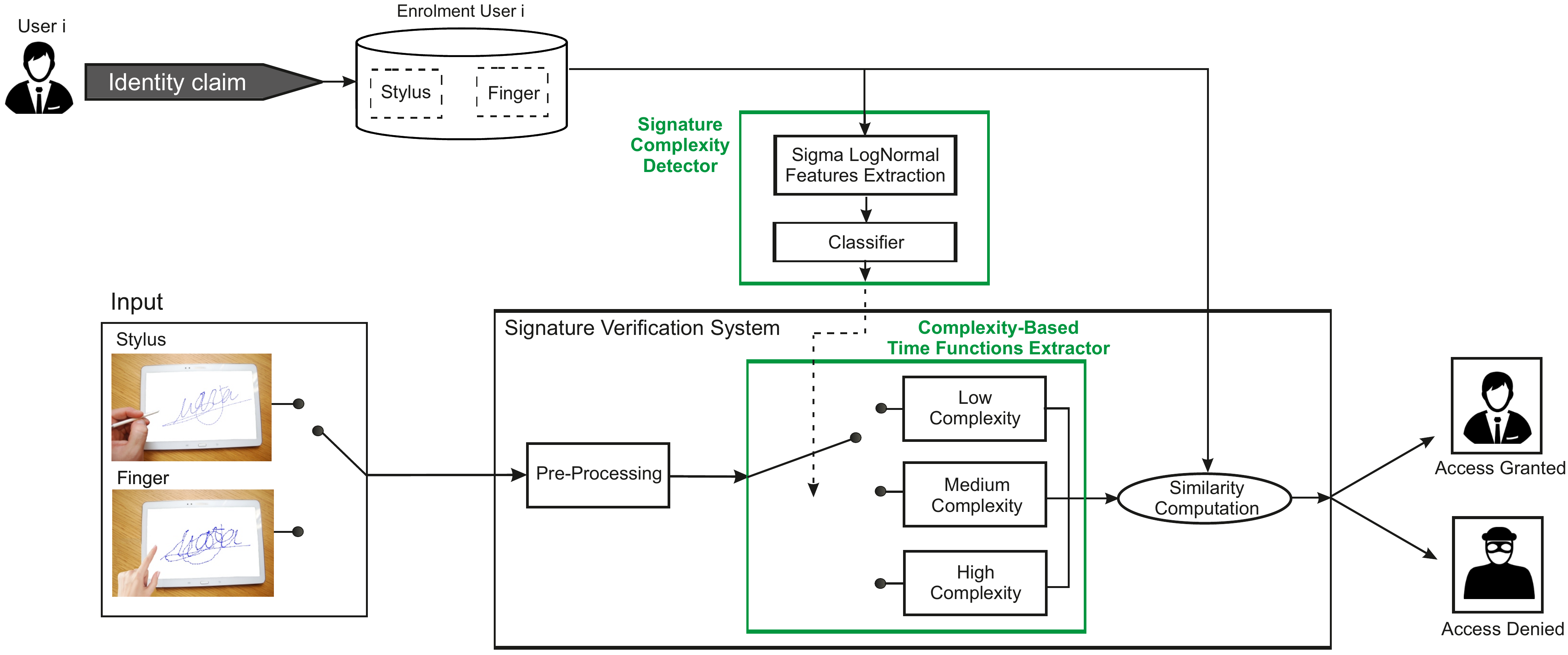}
  \caption{Architecture of our proposed on-line signature verification system adapted to the signature complexity level. Pen, finger, and mixed input scenarios are analysed through the BiosecurID and e-BioSign databases.}
  \label{fig:graphical_abstract}
\end{figure*}

Despite all the improvements achieved in on-line signature verification in recent years, there are still practical challenges that require further research~\cite{Reillo_mobile_2016}. Signatures have been traditionally acquired in pen-based office-like scenarios using devices specifically designed to capture dynamic signatures and handwriting (i.e., so called graphic or writing tablets such as those manufactured by Wacom and others), in which the pen has always been considered as the input device achieving, in general, very good results. However, the high deployment of devices such as smartphones and tablets has given rise to new scenarios where finger and pen are independently considered as input (a.k.a. mixed input). Some preliminary studies have highlighted the challenge of this new scenario~\cite{sae2014online,nam2016mobile,tang2016online,antal2017finger,eBioSign_journal,2018_HanbookBioAntiSpoofing_signature_Tolosana}, but further research is still needed. 

The main goal of this study is to propose a new methodology focused on the development of an on-line signature verification system adapted to the signature complexity level in order to enhance this challenging scenario. Fig. \ref{fig:graphical_abstract} describes our proposed approach based on two main modules: \textit{i)} a signature complexity detector, and \textit{ii)} a complexity-based time functions extractor. Similar approaches have been considered in other biometric traits such as face using the gender and ethnicity information to better train the systems~\cite{FaceGenderID_Vera}, serving as a motivation for our study. A preliminary version of the proposed signature complexity approach was introduced in~\cite{2017_ICDAR_Complexity}. In that study only the pen stylus scenario was considered through the BiosecurID database~\cite{Fierrez2009_PAA}. In general, the present study further analyses the signature complexity effect on both pen and finger scenarios. In addition, we analyse how well our proposed signature complexity approach generalise to other databases and scenarios such as the e-BioSign mobile database.

The main contributions of this study are the following:
\vspace{-4mm}

\begin{itemize}
\item An on-line signature verification system adapted to the signature complexity level is proposed. As far as we know, this is the first study that exploits the signature complexity level to develop more robust and accurate on-line signature verification systems. 
\item A signature complexity detector is proposed. Three different complexity levels are considered regarding the handwriting appearance of the signature, i.e., signatures with an appearance more similar to handwriting are labelled as high complexity whereas signatures with generally simple flourish with no legible information are labelled as low complexity. This simple but effective approach has proven to generalise well to unseen databases and scenarios.
\item We perform an exhaustive comparative analysis between both pen- and touch-based scenarios considering Commercial Off-The-Shelf (COTS) devices and our proposed complexity-based signature verification system.
\end{itemize}

\vspace{-4mm}

The remainder of the paper is organized as follows. In Sec. \ref{sec:related_works}, a review of the most relevant and recent studies related to this work is carried out. In Sec. \ref{sec:proposed_methods}, our proposed complexity-based signature verification system is described. Sec. \ref{sec:signature_database} describes the on-line signature databases considered in the experimental work. Sec. \ref{sec:experimental_work} describes the experimental protocol and the results achieved. Finally, Sec. \ref{sec:conclusions} draws the final conclusions and points out some lines for future work.

\vspace{-7mm}

\section{Related Works}\label{sec:related_works}

\subsection{Signature Complexity}\label{sec:complexity}
Handwritten signature is a biometric trait highly sensitive to the signature complexity~\cite{alewijnse2009analysis,dewhurst2007relationship}. This aspect has been studied in previous studies for both off- and on-line signature verification. 

\subsubsection{Off-Line Signature}\label{sec:off_line_complexity}
In~\cite{Fairhurst1998}, a set of 36 subjects was asked to assign a score based on the visual appearance complexity to five different users whose signatures were of varying length, number of strokes, and with differing degrees of embellishment in signing execution. The results demonstrated that while at the extremes of the scale there is a modest spread in the perceived degree of complexity, the intermediate complexity level appears to be much more difficult to assess and categorise quantitatively. In~\cite{Fernando_Fairhust}, the authors evaluated the effect of complexity and legibility of signatures for off-line signature verification, pointing out the differences in performance for several matchers. 

\subsubsection{On-Line Signature}\label{sec:on_line_signature}
In~\cite{brault1993complexity}, Brault and Plamondon evaluated how signature complexity affects when forging signatures. The authors proposed an automatic difficulty coefficient to measure the difficulty that could be experienced by a typical imitator in reproducing signatures both visually and dynamically. Results obtained using the proposed difficulty coefficient were compared with the opinions of the imitators themselves and an expert document examiner, remarking similarities and differences among them.

A very interesting study was also carried out in~\cite{pepe2012consideration}. Pepe \textit{et al.} analysed the eye movements, pupil changes, and handwriting dynamics while impostors tried to forge two signatures of different complexities. For the experimental framework, 17 subjects participated in the study. A Panasonic NV-GS17 Digital camera was considered to capture the handwriting movements and written trace of the subject while they wrote on a white paper. A PTZ-1230 Wacom Intuos 3 digitizing tablet recorded handwriting dynamics, while a Tobii X-50 eye-tracker simultaneously captured eye movements of subjects. The study concluded with interesting insights: \textit{i)} between complexities, fixations made on the high complexity signature were of greater duration, and \textit{ii)} before the access to the dynamic information of the signatures to forge, 15 of the 17 subjects believed that the high complexity signature would be harder to simulate, however, post-simulation, 12 of the 17 subjects thought that the low complexity signature was harder to simulate. 

Signature complexity has also been associated to the concept of entropy, defining entropy as the inherent information content of biometric samples~\cite{Daugman_iris,lim2015entropy}. In~\cite{Sonia_client_entropy_EER} a ``personal entropy'' measure based on Hidden Markov Models (HMM) was proposed in order to analyse the complexity and variability of on-line signatures regarding three different levels of entropy. Results proved that lower entropy is present in signatures with longer production time and appearance more related to handwriting. In addition, the same authors have proposed a new metric known as ``relative entropy'' for classifying users into animal groups (see the biometric menagerie~\cite{yager2010biometric}) where skilled forgeries are also considered~\cite{Sonia_PlosOne}. 

More recently, Miguel-Hurtado \textit{et al.} proposed in~\cite{miguel2016new} a new approach to automatic signature complexity assessment. They proposed to extract a set of 14 global features such as the total number of \textit{X}-axis intersections and the signature length together with multi-linear regression models to automatically detect the signature complexity level. Their experimental framework was carried out using a private database captured at the University of Kent, comprising 150 participants and using a Wacom Intuos 2 tablet. Their approach achieved a final 78\% success rate. Finally, Sae-Bae \textit{et al.} carried out in~\cite{sae2018distinctiveness} a recent study proposing three different measures to quantify the characteristics of on-line signature templates in terms of distinctiveness, complexity, and repeatability. In particular, the complexity score of a signature is a security measure against skilled forgery attempts. That is, the more complex signature templates are the ones that are harder to forge. The proposed signature complexity score was computed using histogram features based on two factors: \textit{i)} the degree of signature complexity, and \textit{ii)} the inverse of signature template dispersion. Their proposed approach was evaluated for both on-line signature verification and keystroke dynamics, confirming the effectiveness of the approach.

Despite all the studies performed in the on-line signature trait, none of them have exploited the concept of complexity in order to develop better user-adapted on-line signature verification systems, as far as we know. This study intends to further analyse this research line as a novel way to enhance signature verification systems.


\subsection{Pen- and Touch-Based Signature Verification}\label{sec:pen_signature_verification}
The use of the finger as input to signature verification systems has become a thriving scenario for many real commercial applications. However, previous studies in the field have already highlighted how challenging this scenario is for the system performance. In~\cite{marcos13doodles}, both pen and finger were considered as input in the experimental framework. For the finger case, users were asked to perform a simplified version of their signatures (a.k.a. pseudo-signatures) based on their initials or part of their signature flourish. The results using both inputs were analysed, showing a high degradation of the performance for the finger scenario with results in the range of 20.0\% EER. In~\cite{Robertson2015169}, a statistical analysis was conducted to assess consistency between signatures acquired using pen and finger. A set of static and dynamic features that keeps stability in both scenarios was proposed. In~\cite{sae2014online}, the authors acquired a database composed of 6 sessions. Users were asked to perform their signatures using the finger as input on their own devices. Regarding the experimental work, they considered a feature-based system whose features were extracted from histograms related to \textit{X} and \textit{Y} coordinates, speed, angles, pressure, and their derivatives. This approach was evaluated only for random forgeries achieving results between 3.0\% and 8.0\% EERs.

In~\cite{eBioSign_journal}, a benchmark evaluation was reported for the pen, finger, and mixed input scenarios through the e-BioSign database. This database includes dynamic signature and handwriting information acquired using 5 different COTS devices in two separate sessions for a total of 65 users. The results achieved remark the high system performance degradation produced for skilled forgeries when the finger is considered as input with EERs ca. 20.0\%. Nevertheless, for random forgeries, the results provided in that benchmark showed the high feasibility of these new scenarios for real applications with results below 1.0\% EER.


Other studies have also evaluated touch-based signature biometrics on COTS devices. In~\cite{antal2017finger}, both pen and finger scenarios were considered as input. For the pen case, the MCYT database was used whereas for the finger case a new database named MOBISIG was captured using a Nexus 9 tablet with a total of 83 users and 3 acquisition sessions. The results obtained using both feature-based and time functions-based signature verification systems showed the worsening of system performance when the finger is used as input, especially for skilled forgeries with EERs ca. 20.0\%. Similar results have also been obtained in other recent works on the finger scenario using approaches based on autoencoders or simplified versions of Dynamic Time Warping (DTW)~\cite{nam2016mobile,tang2016online}. A very interesting analysis of the mobile finger scenario has been recently presented in~\cite{impedovo2018automatic}. In that study, Impedovo and Pirlo discussed relevant aspects such as accessibility, usability, interoperability, security, and performance. Achievements as well as weakness were discussed in order to highlight promising directions for further research and technology development.

As a conclusion of this section, on-line signature verification systems based on finger input only seem to be feasible in real applications for random forgery impostors. For that scenario, results below 1.0\% EER have been achieved. However, when the expertise of the impostor increases, a high degradation of the system performance is produced with results in the range of 20.0\% EER. 

\section{Proposed Methods}\label{sec:proposed_methods}
Our proposed signature complexity methodology consists of two main modules: \textit{i)} a signature complexity detector, and \textit{ii)} a complexity-based time functions extractor.


\subsection{Signature Complexity Detector}\label{sec:signature_complexity_detector}
The proposed detector classifies each signature into one specific complexity level. Three different complexity levels are proposed regarding the handwriting appearance of the signature: signatures with an appearance more similar to handwriting are labelled as high complexity whereas signatures with generally simple flourish with no legible information are labelled as low complexity. We propose to use the number of strokes as a simple measure of the signature complexity. In this study we extract this information through the well-known writing generation Sigma LogNormal model, which was first introduced to on-line signature in~\cite{Plamondon_sigmaLogNormal}. This model has been widely used in many different tasks such as signature verification~\cite{2015_ICB_skilledSignSigmaLog_Marta,Plamondon_Fisher}, recovering on-line signatures from image-based specimens~\cite{diaz2017recovering}, and to monitor a range of neuromuscular diseases~\cite{impedovo2018dynamic}, among many others. In particular, we consider in this study the popular ScriptStudio public software provided by the authors. We would like to highlight that Ferrer \textit{et al.} has recently proposed in~\cite{ferrer2018idelog} a novel framework named iDeLog. This novel approach is able to reconstruct the trajectory significantly better than ScriptStudio when movement is continuous, long, and complex. However, iDeLog provides worse results than ScriptStudio in reconstructing velocity, due to the trade-off between trajectory and velocity in iDeLog.  

The Sigma LogNormal model emulates the physiological human movement production for the generation of signatures. The idea is based on the fact that one signature can be decomposed into strokes in which each stroke $i$ follows a lognormal velocity distribution $\vec{v_i}(t)$:
\vspace{-8mm}

\begin{equation}
|\vec{v_i}(t)|=\frac{D_i}{\sigma_i(t-t_{0{i}})\sqrt{2\pi}}\exp \left(\frac{(\ln (t-t_{0{i}})- \mu_i)^{2}}{-2\sigma_i^{2}}\right)
\label{ecuacion1}
\end{equation}

\noindent{where $t_{0{i}}$ is the starting time of the stroke, $D_i$ its length, $\mu_i$ the log time delay and $\sigma_i$ the log response time. In addition, the angular position of each stroke along a pivot direction is expressed through the start angle $\theta_s$ and the end angle $\theta_e$. Thus, each stroke is represented by ($D_i$, $t_{0{i}}$, $\mu_i$, $\sigma_i$, $\theta_{si}$, $\theta_{ei}$). The complete velocity profile of one signature can be modelled as a sum of the different individual stroke velocity profiles as:}

\begin{equation}
\vec{v}(t) = \sum_{i=1}^{N} \vec{v_i}(t)   
\end{equation}     

\begin{figure}[tb]
\centering
\subfigure{\label{figGenPen}
\includegraphics[width=0.8\linewidth]{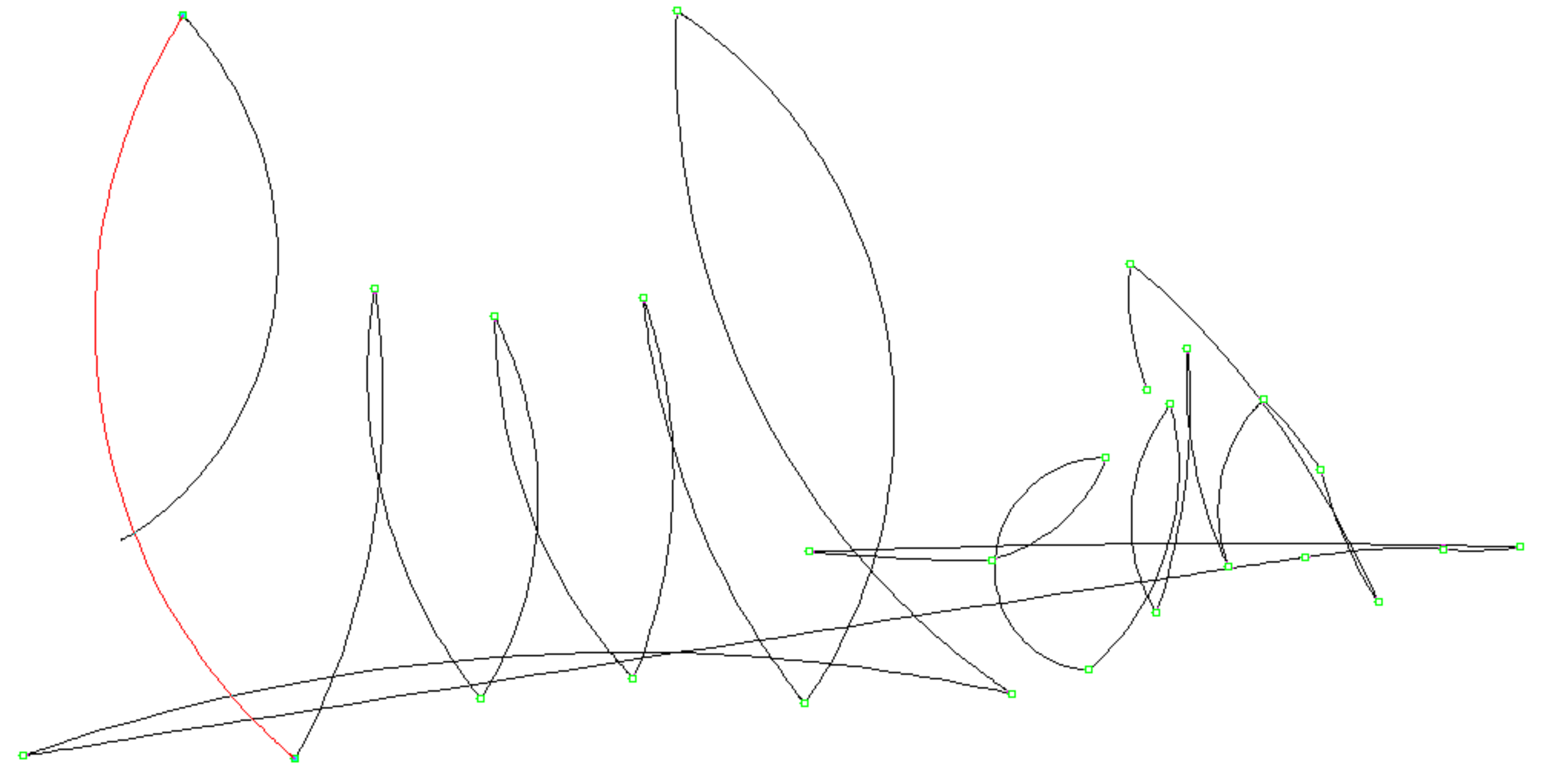}}
\vspace{0.2cm}
\subfigure{\label{figGenPen}
\includegraphics[width=0.8\linewidth]{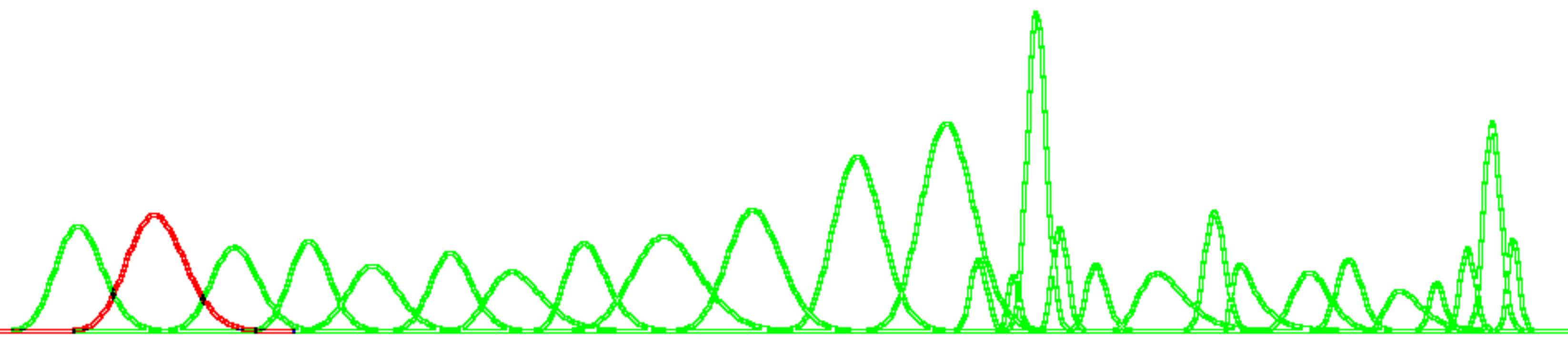}}
\caption{Trace and velocity profile of one reconstructed on-line signature using the Sigma LogNormal model. A single stroke of the signature and its corresponding lognormal profile are highlighted in red colour. Individual strokes are segmented within the LogNormal algorithm~\cite{Plamondon_sigmaLogNormal}.}
 \label{fig_stroke_lognormal}
\end{figure}

\noindent{where $N$ represents the number of strokes involved in the generation of a given signature. Fig. \ref{fig_stroke_lognormal} shows the lognormal velocity profiles extracted for each stroke of one example signature using ScriptStudio. Overall, an average Signal-to-Noise Ratio (SNR) value of around 25 dB is achieved in the databases considered in this study.
}

We propose to use the number of lognormals ($N$) that models each signature as a measure of the complexity level of the signature. It is worth noting that lognormals with amplitude values lower than a threshold were discarded in order to consider only important lognormals directly related to the main strokes performed while signing. Once this parameter is extracted for all available enrolment signatures of a particular user, that user is classified into a complexity level using the majority voting algorithm (the signature complexity level of the majority of the enrolment signatures of that user). At the test stage, we consider the complexity level of the claimed user (see Fig. \ref{fig:graphical_abstract}). In the case that there is no claimed identity, e.g., in signature identification, the complexity level of the identity being compared with the test signature would be used. The advantage of this approach is that the signature complexity detector can be trained and developed as a previous off-line process, e.g., after the enrolment of the user. Therefore, at the verification stage, the complexity of the claimed user is already known, avoiding time consuming delays and making it feasible to be applied in real-time scenarios.

\subsection{Complexity-Based Time Functions Extractor}\label{sec:client_entropy_system}

Once the user is classified into a complexity level, we propose to extract the optimal time functions associated to each specific complexity level (see Fig. \ref{fig:graphical_abstract}). For each signature acquired using the pen or the finger, signals related to \textit{X} and \textit{Y} spatial coordinates are used to extract an initial set of 21 time functions (see Table \ref{tabla:tablaLocalFeatures}). For more details about the time functions implementation, we refer the reader to~\cite{Marcos_features}. In addition, the same approach proposed in~\cite{2015_IEEEAccess_InterSign_Tolosana} is considered in this study in order to mitigate the degradation performance for mixed input scenarios. This approach comprises two main stages:

\begin{itemize}
\item A data preprocessing stage is applied with the aim of obtaining signatures with the same type of information and time and spatial position, regardless of the writing input. Normalisation based on the mean and standard deviation is applied to all signatures for that purpose. In addition, information related to the pressure on the writing surface and pen-up trajectories is removed from those signatures acquired with the pen as this information is not available when the finger is considered as input. Finally, an additional interpolation step based on splines is applied to standardise the sampling frequency to 200 Hz among different input scenarios.

\item A selection of robust and stable time functions coming from the same or different writing input. The Sequential Forward Feature Selection (SFFS) algorithm is considered to select the optimal subset of time functions for each complexity level in terms of EER.
\end{itemize}

\begin{table}[tb]
\centering
\caption{Set of time functions considered in this work. For more details, we \\ refer the reader to~\cite{Marcos_features}.}
\vspace{3mm}
\begin{tabular}{p{1cm} p{7cm}}
\hline
\# & Time Function \\
\hline \hline
1 & \textit{X}-coordinate: $x_n$  \\
\hline
2 & \textit{Y}-coordinate: $y_n$  \\
\hline
3 & Path-tangent angle: $\theta_n$  \\
\hline
4 & Path velocity magnitude: $v_n$ \\
\hline
5 & Log curvature radius: $\rho_n$ \\
\hline
6 & Total acceleration magnitude: $a_n$ \\
\hline
7-12 & First-order derivate of features 1-7: $\dot{x_n},\dot{y_n},\dot{\theta_n},\dot{v_n},\dot{\rho_n},\dot{a_n}$ \\
\hline
13-14 & Second-order derivate of features 1-2: $\ddot{x_n},\ddot{y_n}$ \\
\hline
15 & Ratio of the minimum over the maximum speed over a 5-samples window: $v^r_n$ \\
\hline
16-17 & Angle of consecutive samples and first order difference: $\alpha_n$, $\dot{\alpha_n}$ \\
\hline
18 & Sine of the angle of consecutive samples: $s_n$ \\
\hline
19 & Cosine of the angle of consecutive samples: $c_n$ \\
\hline
20 & Stroke length to width ratio over a 5-samples window: $r^5_n$ \\
\hline
21 & Stroke length to width ratio over a 7-samples window: $r^7_n$ \\
\hline
\end{tabular}
\label{tabla:tablaLocalFeatures}
\end{table}



\section{On-Line Signature Databases}\label{sec:signature_database}

Two different public databases are considered in the experimental framework of this study~\footnote{https://github.com/BiDAlab/DeepSignDB}. 

\subsection{e-BioSign}\label{sec:e_BioSign}
For the e-BioSign database~\cite{eBioSign_journal}, we consider a subset of the full database composed of signatures acquired using a Samsung ATIV 7 general purpose device (a.k.a. W4 device). The W4 device has a 11.6-inch LED display with a resolution of 1920$\times$1080 pixels and 1024 pressure levels. Data was collected using a pen stylus and also the finger in order to study the performance of signature verification in a mobile scenario. The available information when using the pen stylus is \textit{X} and \textit{Y} pen coordinates and pressure. In addition, pen-up trajectories are also available. However, pressure information and pen-ups trajectories are not recorded when the finger is used as input. Regarding the acquisition protocol, the device was placed on a desktop and subjects were able to rotate the device in order to feel comfortable with the writing position.

Data was collected in two sessions for 65 subjects with a time gap between sessions of at least 3 weeks. For each user and writing input, there are a total of 8 genuine signatures and 6 skilled forgeries. It is important to note the high quality of skilled forgeries for both pen and finger inputs as forgers had access to the dynamic realization of the signatures to be forged. 

\subsection{BiosecurID}\label{sec:e_BioSign}
For the BiosecurID database~\cite{Fierrez2009_PAA}, signatures were acquired from 400 users through a Wacom Intuos 3 pen tablet with a resolution of 5080 dpi and 1024 pressure levels. The database comprises a total of 16 genuine signatures and 12 skilled forgeries per user, captured in 4 separate acquisition sessions leaving a two-month interval between them, and in a controlled and supervised office-like scenario. Signatures were acquired using only a pen stylus. Information related to \textit{X} and \textit{Y} pen coordinates, pressure, and pen-up trajectories is available for each signature.

\section{Experimental Framework}\label{sec:experimental_work}

\subsection{Signature Verification Matcher}\label{sec:experimental_protocol}
The popular DTW algorithm is used to compute the similarity between the time functions from the input and training signatures of the claimed user. In particular, we consider the implementation proposed in~\cite{Marcos_matching}. For the computation of the distance measure between sequence samples, we use Euclidean distance. For the definition of the weighting factors, only three transitions with the same value equal to 1 are allowed for the computation of the accumulated distance, which is finally normalised by the length of the warping path. Once we have the accumulated distance $D$, the similarity computation score $s$ is obtained as $s = \exp(-D)$.

It is important to remark that the same DTW scheme is always considered for obtaining the similarity score regardless of the signature complexity level.

\subsection{Experimental Protocol}\label{sec:experimental_protocol}
The experimental protocol is designed to provide a fair evaluation of both the signature complexity detector and the complexity-based time functions extractor on pen and finger scenarios. Both BiosecurID and e-BioSign databases are divided into development (40\% of the users) and evaluation (60\% of the remaining users). 

For the evaluation of each module, the 4 genuine signatures of the first session are used as reference signatures, whereas the remaining genuine signatures (i.e., 4 and 12 for the e-BioSign and BiosecurID databases, respectively) are used for testing. Skilled forgery scores are obtained by comparing the reference signatures against the available skilled forgeries of each user (i.e., 6 and 12 for the e-BioSign and BiosecurID databases, respectively) whereas random (zero-effort) forgery scores are obtained by comparing the reference signatures with one genuine signature of each of the remaining users of the same database. The final score is obtained after performing the average score of the four one-to-one comparisons.



Finally, the following nomenclature is used for the different input scenarios considered: ``training-testing'', where ``training'' and ``testing'' mean the writing tool considered for the training and testing signatures, respectively. For example, the case ``pen-finger'' means that signatures considered for training are acquired using the pen whereas signatures considered for testing are acquired with the finger.

\section{Experimental Results}\label{sec:experimental_results}

\subsection{Signature Complexity Detector}\label{sec:entropy}

\begin{figure}[tb]
  \centering
    \includegraphics[width=\linewidth]{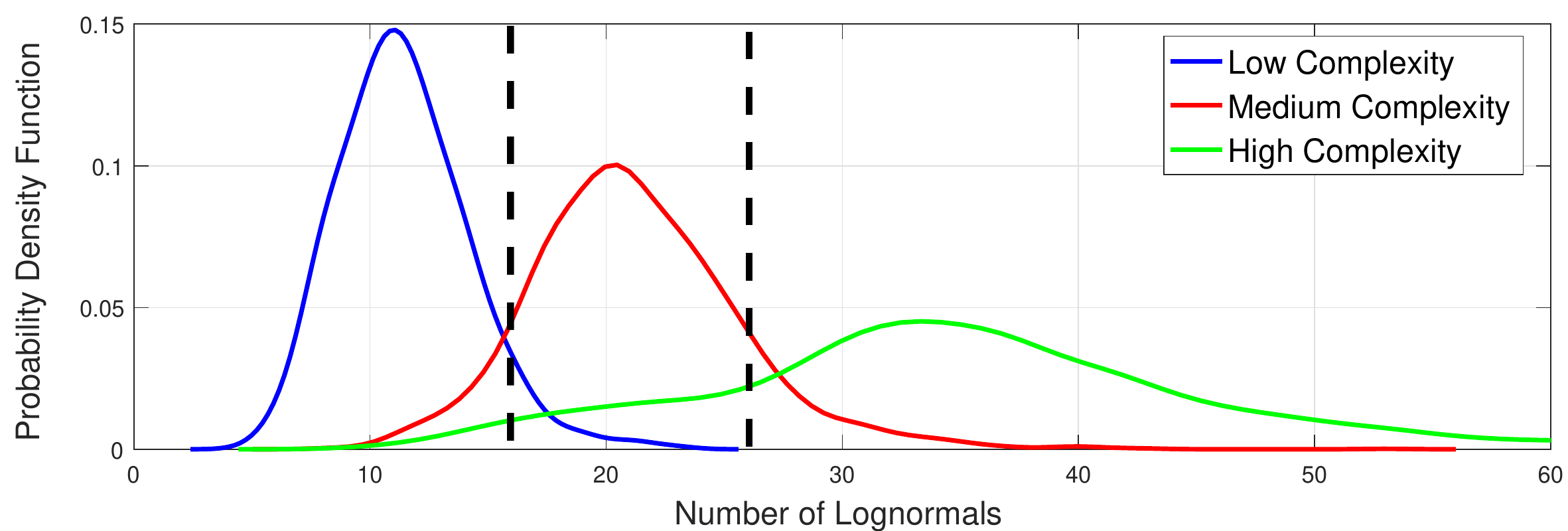}
  \caption{Probability density function of the number of lognormals for each manually annotated complexity level using all genuine signatures of the BiosecurID database. The three proposed complexity-dependent decision thresholds are highlighted by black dashed lines.}
  \label{fig:all_sigma_logNormals}
\end{figure}
\begin{figure}[tb]
     \centering
     \subfigure[BiosecurID]{\label{fig:BiosecurID_database}
     \includegraphics[width=\linewidth]{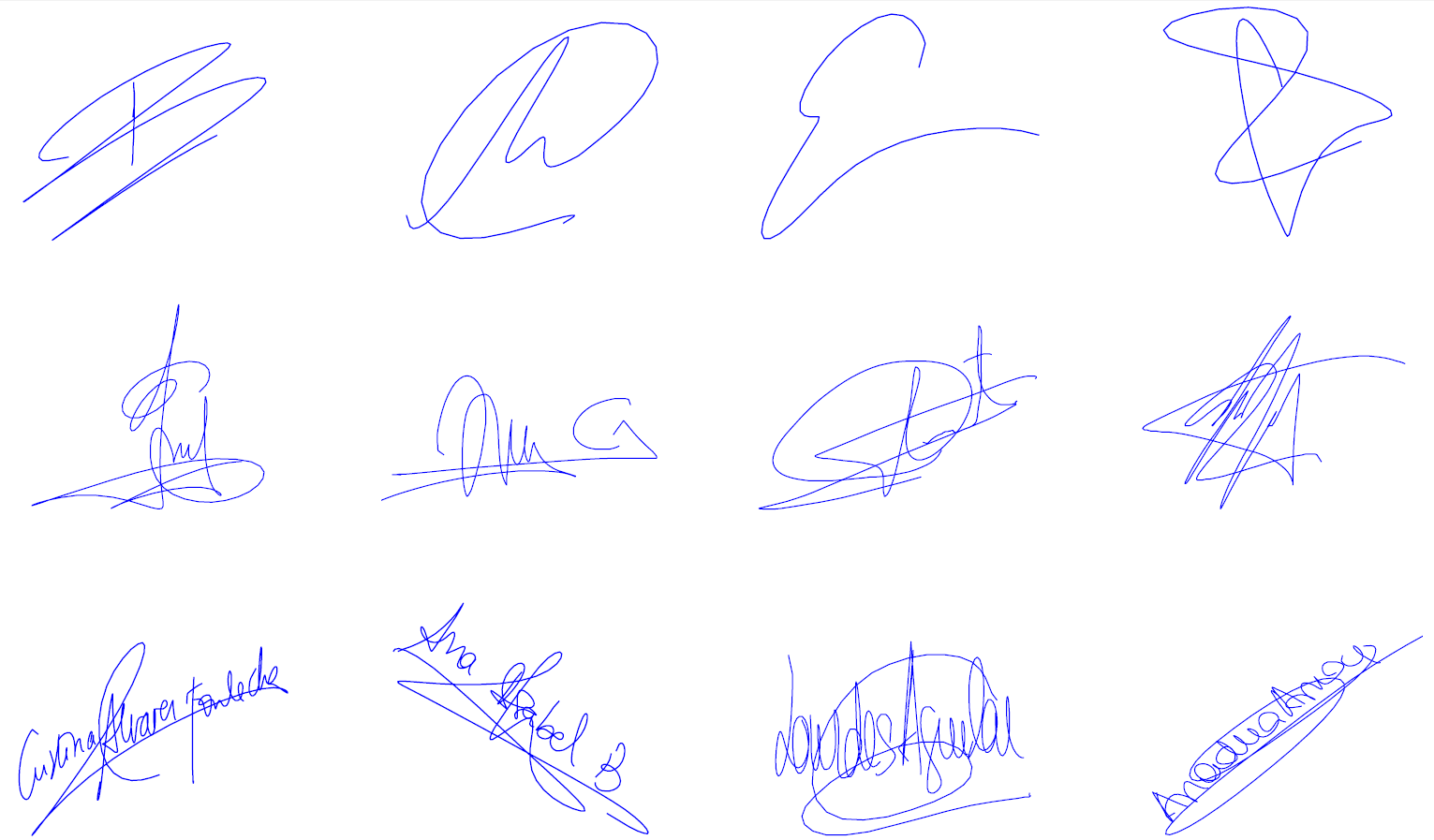}}
      \subfigure[e-BioSign]{\label{fig:eBioSign_database}
     \includegraphics[width=\linewidth]{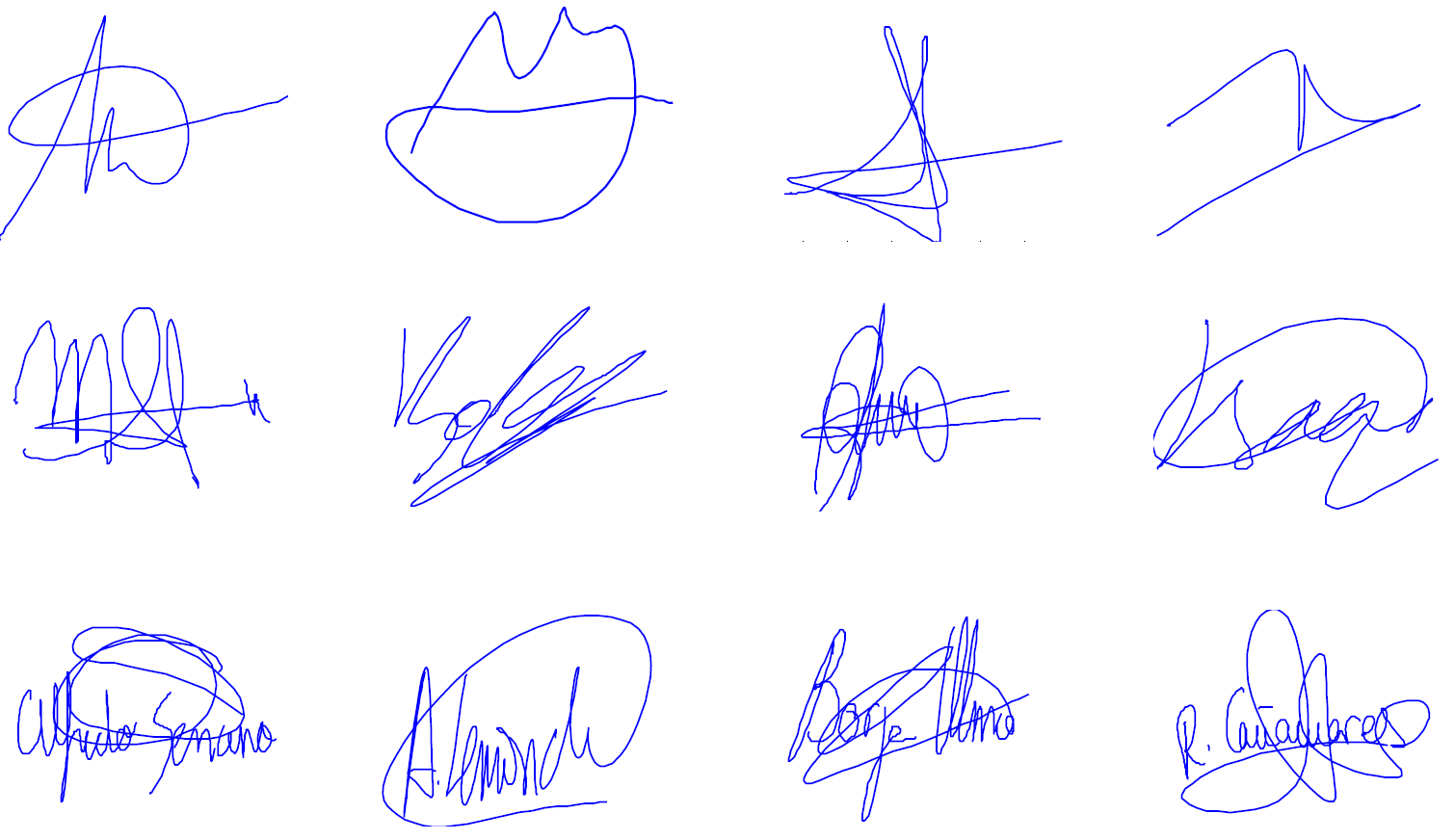}}
     \caption{Signatures categorised into each complexity level using our proposed signature complexity detector. From top to bottom: low, medium, and high complexity.}
     \label{fig:signatures_entropy}
\end{figure}

The signature complexity detector was developed in two different steps. First, each user of the BiosecurID database was manually labelled in a signature complexity level (i.e., low, medium, and high) based on previous studies~\cite{Sonia_PlosOne}. This process was carried out by two different annotators twice each in order to keep consistency on the results. The image of just one genuine signature per user was visualised to classify each user into a complexity label. Users whose signatures are with an appearance more similar to handwriting were labelled as high-complexity users whereas those users whose signatures are generally simple flourish with no legible information were labelled as low-complexity users. This first stage served as a ground truth. Then, we extracted the number of lognormals $N$ for each available genuine signature of the BiosecurID database (i.e., a total of $400\times16 = 6400$ genuine signatures). Following this stage, we represented for each complexity level its corresponding distribution of lognormals according to the ground truth generated during the first stage. Fig. \ref{fig:all_sigma_logNormals} shows the distributions of the number of lognormals obtained for each complexity level using all genuine signatures of the BiosecurID database. The three proposed complexity-dependent decision thresholds are highlighted by black dashed lines. They were selected in order to minimize the number of misclassifications between different signature complexity levels. Signatures with lognormal values equal or less than 17 are classified as low-complexity signatures whereas those signatures with more than 27 lognormals are classified into the high-complexity group. Otherwise, signatures are categorised into the medium-complexity level. Additionally, an analysis of the stability regarding the number of lognormals for different signatures of the same user is carried out in order to assess the feasibility of our proposed signature complexity detector. In general, low standard deviation values are obtained. Users with a low-complexity level provide an average number of 12.5 lognormals and a standard deviation of 1.3 whereas medium- and high-complexity levels achieve averages of 21.1 and 31.3 lognormals with standard deviations of 2.6 and 3.9, respectively. These results make sense as the intra-user variability increases with the signature complexity level. The same thresholds are extrapolated to the e-BioSign database in order to study the generalisation capacity of the proposed approach to unseen databases and scenarios. Fig. \ref{fig:signatures_entropy} depicts some of the signatures classified into each complexity level for both BiosecurID and e-BioSign databases using our proposed approach.


\begin{table*}[tb]
\centering
\caption{\textbf{Signature complexity detector:} Signature verification performance in terms of EER (\%) for each complexity level using the pen evaluation datasets of BiosecurID and e-BioSign. Skilled and random forgery results are shown on top and bottom of each cell respectively.}
\vspace{3mm}
\label{detector_BiosecurID_eBioSign}
\begin{tabular}{c c c c}
\multicolumn{1}{c}{} & \textbf{Low Complexity}                                             & \textbf{Medium Complexity}                                           & \textbf{High Complexity}                                              \\ \hline
\multicolumn{1}{c}{BiosecurID} & \begin{tabular}[c]{@{}c@{}}13.8\\ 1.5\end{tabular} & \begin{tabular}[c]{@{}c@{}}7.5\\ 0.7\end{tabular} & \begin{tabular}[c]{@{}c@{}}6.2\\ 0.9\end{tabular} \\ \hline
\multicolumn{1}{c}{e-BioSign} & \begin{tabular}[c]{@{}c@{}}11.1\\ 0.1\end{tabular} & \begin{tabular}[c]{@{}c@{}}8.3\\ 0.1\end{tabular} & \begin{tabular}[c]{@{}c@{}}5.6\\ 0.1\end{tabular} \\ \hline
\end{tabular}
\end{table*}

We now evaluate our proposed signature complexity detector following the same procedure carried out in~\cite{Sonia_PlosOne}: analysing the system performance of each different complexity group considering state-of-the-art signature verification systems as Baseline Systems~\cite{eBioSign_journal,2015_ICB_skilledSignSigmaLog_Marta}. These Baseline Systems are based on DTW and a selection of the most discriminative time functions for each database through the SFFS algorithm, regardless of the complexity level.

Table \ref{detector_BiosecurID_eBioSign} shows the system performance results in terms of EER(\%) for each complexity level considering the pen evaluation datasets. Each user is classified into a complexity level applying the majority voting algorithm to the 4 enrolment signatures of the user.


Results highlight the different signature verification performance regarding the signature complexity level. Users with a high-complexity level achieve an absolute improvement of 7.6\% and 5.5\% EER compared with the users categorised into a low-complexity level for the BiosecurID and e-BioSign databases, respectively. Similar results were obtained in previous studies using other approaches~\cite{Sonia_PlosOne}. In that work, users categorised into a high-complexity level achieved an absolute improvement of 8.5\% EER compared with users categorised into a low-complexity level for the MCYT database. These results prove the effectiveness and generalisation of the proposed signature complexity detector to other databases and scenarios.

In the following sections we analyse the idea of considering an on-line signature verification system adapted to the signature complexity level so as to further reduce the system performance.

\subsection{Complexity-Based Time Functions Extractor}\label{sec:time_functions_selection}

\subsubsection{Time Functions Selection}\label{sec:selection}
This section analyses which are the most discriminative and robust time functions for each signature complexity level using the SFFS algorithm over the development datasets. For the BiosecurID database, 4 and 12 genuine signatures from the first and remaining available sessions are considered as training and testing signatures, respectively. For the e-BioSign database, a total of 4 genuine signatures from the first session (2 signatures per writing input) and 8 genuine signatures from the second session (4 signatures per writing input) are considered as training and testing signatures, respectively. A separate optimal time-function vector is extracted for each complexity level regardless of the writing input used while signing.


The following three cases are studied:

\begin{itemize}
\item Time functions selected for all three signature complexity levels, i.e., complexity-independent: CX-All.
\item Time functions selected for only medium and high signature complexity: CX-High.
\item Time functions selected for only low and medium signature complexity: CX-Low.
\end{itemize}

Table \ref{best_timeFunctions} shows the time functions automatically selected for each considered case. Different sets of time functions result for the BiosecurID and e-BioSign databases as the former considers an office-like scenario whereas the latter considers a mobile scenario. For BiosecurID, the time functions $\dot{a_n}$ and $v^r_n$ are selected in CX-all, whereas for e-BioSign the time functions are $y_n$ and $\dot{x_n}$. In BiosecurID the time functions are more related to the acceleration and speed of the users performing their signatures, whereas in e-BioSign the functions related to the position of the writing tool (i.e., \textit{X} and \textit{Y} pen coordinates) are more adequate in CX-all. The reason behind this difference seems to be the mixed input scenario of e-BioSign. 

In CX-high, very similar time functions are selected for BiosecurID and e-BioSign. These time functions provide information related to the variation of the velocity, vertical acceleration and variation of angle, time functions more related to the geometry of characters and therefore to handwriting.

Finally, time functions such as $c_n$ and $s_n$ are selected in CX-low, providing information related to the signature trajectory angles, as expected for simple signatures with no legible information. 

\begin{table}[tb]
\centering
\caption{Time functions selected for different complexities (CX) and databases.}
\vspace{3mm}
\label{best_timeFunctions}
\begin{tabular}{c c c c}
\multicolumn{1}{c}{} & \textbf{CX-All} & \textbf{CX-High} & \textbf{CX-Low} \\ \hline
\multicolumn{1}{c}{BiosecurID} & $\dot{a_n}$, $v^r_n$        & $\dot{v_n}$, $\ddot{y_n}$, $\dot{\alpha_n}$        & $c_n$       \\ \hline
\multicolumn{1}{c}{e-BioSign}  & $y_n$, $\dot{x_n}$       & $\dot{\theta_n}$, $\dot{y_n}$, $v^r_n$        & $x_n$, $s_n$        \\ \hline
\end{tabular}
\end{table}


\subsubsection{Pen Scenario}\label{sec:pen_scenario}
This section evaluates our proposed complexity approach for the case of using the pen stylus both for training and testing (i.e., Pen-Pen). Table \ref{pen_scenarios} shows the results achieved for both BiosecurID and e-BioSign evaluation datasets. The same Baseline System described and used in Sec. \ref{sec:entropy} are considered here in order to measure the improvements achieved by our proposed approach. It is important to remark that the only two differences between the Proposed and Baseline Systems are: \textit{i)} the proposed signature complexity detector, and \textit{ii)} the proposed complexity-based time functions extractor. Thus, the same DTW scheme is always considered for obtaining the similarity scores.

\begin{table*}[tb]
\centering

\caption{\textbf{Pen scenario:} System performance results in terms of EER (\%) of each complexity level using the evaluation datasets of BiosecurID and e-BioSign. Skilled and random forgery results are shown on top and bottom of each cell respectively.}
\vspace{3mm}
\label{pen_scenarios}
\begin{tabular}{c c c| c c| c c }
\multicolumn{1}{c|}{\textbf{Dataset (Training-Testing)}} & \multicolumn{2}{c|}{\textbf{Low Complexity}} & \multicolumn{2}{c|}{\textbf{Medium Complexity}}                                                         & \multicolumn{2}{c}{\textbf{High Complexity}}  \\  \multicolumn{1}{c|}{}               & \textbf{Baseline}                                           & \textbf{Proposed}                                           & \textbf{Baseline}                                           & \textbf{Proposed}                                           & \textbf{Baseline}                                           & \textbf{Proposed}                                           \\ \hline
\multicolumn{1}{c|}{BiosecurID (Pen-Pen)} & \begin{tabular}[c]{@{}c@{}}13.8\\ 1.5\end{tabular} & \begin{tabular}[c]{@{}c@{}}10.1\\ 1.3\end{tabular}  & \begin{tabular}[c]{@{}c@{}}7.5\\ 0.7\end{tabular}  & \begin{tabular}[c]{@{}c@{}}5.2\\ 0.5\end{tabular} & \begin{tabular}[c]{@{}c@{}}6.2\\ 0.9\end{tabular}  & \begin{tabular}[c]{@{}c@{}}4.6\\ 0.9\end{tabular} \\ \hline
\multicolumn{1}{c|}{e-BioSign (Pen-Pen)} & \begin{tabular}[c]{@{}c@{}}11.1\\ 0.1\end{tabular} & \begin{tabular}[c]{@{}c@{}}8.3\\ 0.1\end{tabular}  & \begin{tabular}[c]{@{}c@{}}8.3\\ 0.1\end{tabular}  & \begin{tabular}[c]{@{}c@{}}10.2\\ 0.1\end{tabular} & \begin{tabular}[c]{@{}c@{}}5.6\\ 0.1\end{tabular}  & \begin{tabular}[c]{@{}c@{}}5.6\\ 0.1\end{tabular} \\ \hline
\end{tabular}
\end{table*}

Analysing the skilled forgery results obtained for the BiosecurID database, our Proposed System achieves an average absolute improvement of 2.5\% EER compared with the Baseline System. It is important to remark that for the most challenging users (i.e., users with a low-complexity level as they are easier to forge), our proposed approach achieves an absolute improvement of 3.7\% EER compared with the Baseline System. Analysing the results obtained for random forgeries, our Proposed System also achieves improvements. This improvement is lower compared with the skilled forgery scenario as the SFFS algorithm is focused on the skilled forgeries, not random.

Regarding the e-BioSign database, our Proposed System also achieves similar trends. The improvement is slightly lower compared with the BiosecurID.

\begin{figure}[tb]
     \centering
     \subfigure[BiosecurID]{\label{fig:BiosecurID_pen}
     \includegraphics[width=0.6\linewidth]{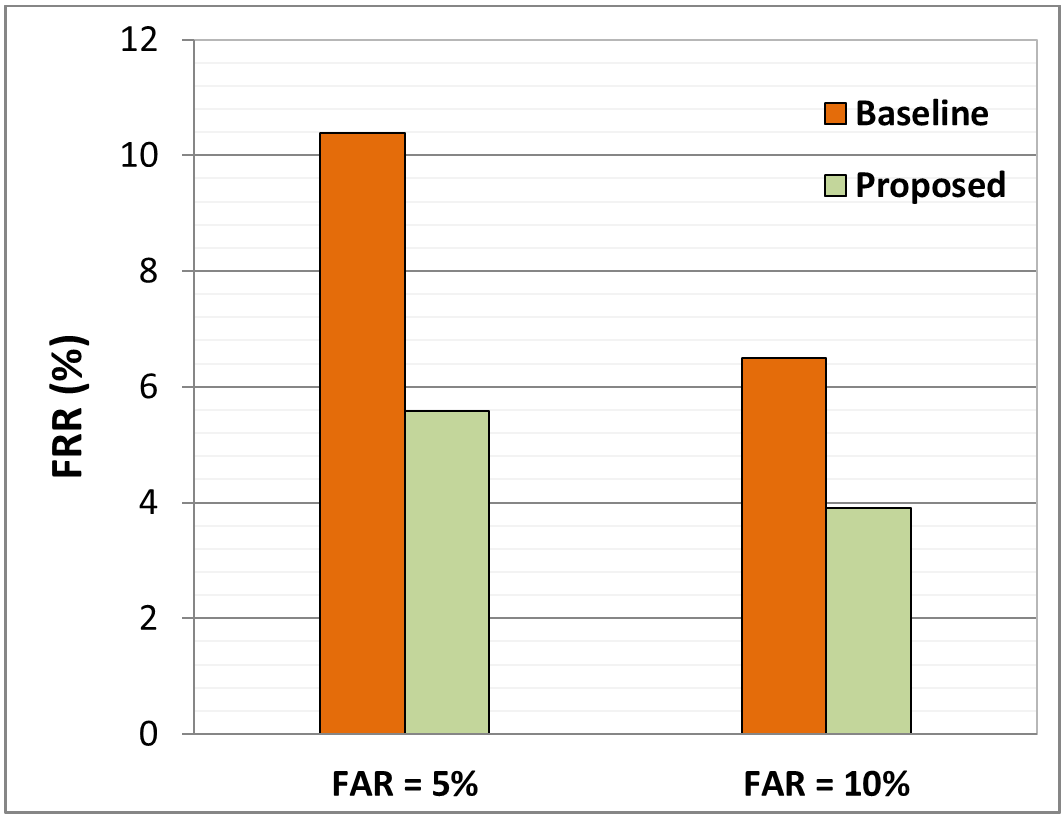}}
      \subfigure[e-BioSign]{\label{fig:eBioSign_pen}
     \includegraphics[width=0.6\linewidth]{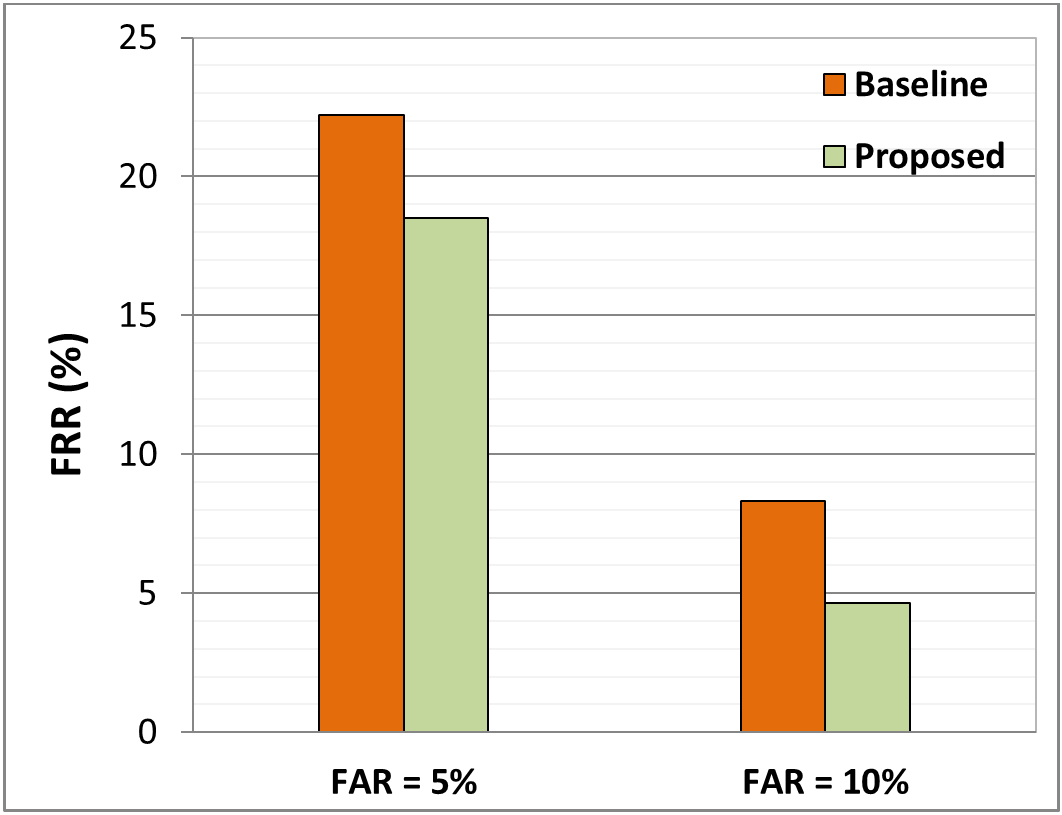}}
     \caption{\textbf{Pen scenario:} system performance results in terms of FRR at different values of FAR using the evaluation datasets.}
     \label{fig:pen_BiosecurID_eBioSign}
\end{figure}

Finally, we provide in Fig. \ref{fig:pen_BiosecurID_eBioSign} the system performance results in terms of the False Rejection Rate (FRR) at different values of False Acceptance Rate (FAR) for both Baseline and Proposed Systems considering all complexity levels together. We consider this visualisation approach, and not the traditional Detection Error Tradeoff (DET) curves, as different system thresholds are considered regarding the complexity level. Our Proposed System achieves an average absolute improvement of 3.7\% FRR. In particular, for a FAR = 10\%, final values of 3.9\% and 4.6\% FRR are achieved for BiosecurID and e-BioSign, respectively. These results show the improvement of on-line signature verification systems adapted to the signature complexity level. 

\begin{table*}[tb]
\centering
\caption{\textbf{Pen scenario:} Comparison of our proposed approach with previous studies in terms of FAR and FRR for the BiosecurID database. The number of users included on each complexity level has been computed by the signature complexity detector proposed in this work.}
\vspace{3mm}
\label{comparison_otherWorks}
\begin{adjustbox}{width=0.97\textwidth}
\begin{tabular}{c| c| c c c c c c c}
\hline
\textbf{Work}                       & \textbf{Algorithm}                                                                            & \begin{tabular}[c]{@{}c@{}}\textbf{Inter-Session} \\ \textbf{Variability}\end{tabular} & \begin{tabular}[c]{@{}c@{}}\# \textbf{Training}\\ \textbf{Signatures}\end{tabular} & \begin{tabular}[c]{@{}c@{}}\textbf{\% Users}\\  \textbf{CX-Low}\end{tabular} & \begin{tabular}[c]{@{}c@{}}\textbf{\% Users} \\ \textbf{CX-Medium}\end{tabular} & \begin{tabular}[c]{@{}c@{}}\textbf{\% Users}\\ \textbf{CX-High}\end{tabular} & \textbf{FAR (\%)}          & \textbf{FRR (\%)}          \\ \hline
Ferrer \textit{et al.}~\cite{ferrer2016behavioral}                     & DTW & No                                                                   & 5                                                                & 7.6                                                        & 39.4                                                          & 53.0                                                       & 3.1          & 3.1 \\ \hline 
Diaz \textit{et al.}~\cite{diaz2017recovering}                    & Manhattan Distance                                                                       & No                                                                   & 5                                                                & 8.0                                                        & 36.0                                                          & 56.0                                                       & 3.2          & 3.2          \\  \hline  
Galbally \textit{et al.}~\cite{galbally2015line}                   & DTW & Yes                                                                  & 4                                                                & 7.6                                                        & 39.4                                                          & 53.0                                                       & 6.9          & 6.9          \\  \hline
Gomez-Barrero \textit{et al.}~\cite{2015_ICB_skilledSignSigmaLog_Marta}                      & \begin{tabular}[c]{@{}c@{}}Impostor Detector\\ + DTW\end{tabular}            & Yes                                                                  & 4                                                                & 7.2                                                        & 38.0                                                          & 54.8                                                       & 4.8          & 4.8          \\ \hline \hline
Baseline System            & DTW & Yes                                                                  & 4                                                                & 9.6                                                        & 37.9                                                          & 52.5                                                       & 5.0          & 10.4         \\ \hline 
\textbf{Proposed Approach} & \textbf{\begin{tabular}[c]{@{}c@{}}Complexity-based\\ DTW\end{tabular}} & \textbf{Yes}                                                         & \textbf{4}                                                       & \textbf{9.6}                                               & \textbf{37.9}                                                 & \textbf{52.5}                                              & \textbf{5.0} & \textbf{5.8} \\ \hline
\end{tabular}
\end{adjustbox}
\end{table*}

We now compare our proposed complexity-based signature verification system with other existing state-of-the-art approaches that have been evaluated using the BiosecurID database. The comparison is not straightforward as different experimental protocols are considered in each of the studies. This is something worth highlighting, not only for this comparison, but also for the future of the field as results can vary significantly depending on the particular protocol used. For this reason, in order to perform a fair comparison with other studies, Table \ref{comparison_otherWorks} depicts not only the FAR and FRR values achieved for each approach but also other very important aspects that affect the final system performance such as the complexity level of the evaluated users and the inter-session signature comparisons when testing. Our Proposed System outperforms the results achieved in our previous study~\cite{galbally2015line}, where the same DTW scheme was considered but not the complexity concept for the time functions extraction. Besides, very similar results are achieved compared with~\cite{2015_ICB_skilledSignSigmaLog_Marta}, in which a skilled forgery detector was incorporated to an already competitive DTW system. Finally, our proposed approach is also compared with other approaches based on Manhattan distance~\cite{diaz2017recovering}, producing worse results due to a different number of training signatures, percentages of users in the complexity levels, and mainly due to not considering inter-session signature comparisons when testing. This critical effect can be observed in~\cite{ferrer2016behavioral} as well, where better results are achieved when applying a simple DTW approach based only on \textit{X} and \textit{Y} coordinates and their derivatives.

\subsubsection{Finger vs. Pen Scenarios}\label{sec:finger_mixed_scenarioos}

This section evaluates our proposed complexity approach for the case of using the finger both for training and testing (i.e., Finger-Finger). Therefore, only the e-BioSign evaluation dataset is used in this section as signatures acquired using the finger are not available in the BiosecurID database. The same Baseline and Proposed Systems considered in the previous section are analysed here.

First, we analyse the results obtained for the finger input scenario (i.e., Finger-Finger). Table \ref{finger_mixed_scenarios} depicts the system performance results in terms of EER (\%) of each complexity level using the evaluation dataset of e-BioSign. Analysing the skilled forgery results, our Proposed System achieves an average absolute improvement of 3.4\% EER compared with the Baseline System. Similar to the pen scenario, the highest improvement is achieved for the most challenging users (i.e., users with a low-complexity level) with an absolute improvement of 5.6\% EER compared with the Baseline System. Regarding random forgeries, the same very good results (close to 0.0\% EER) are achieved using our proposed approach.

\begin{table*}[tb]
\centering
\caption{\textbf{Pen, finger, and mixed input scenarios:} System performance results in terms of EER (\%) of each complexity level using the evaluation dataset of e-BioSign. Skilled and random forgery results are shown on top and bottom of each cell respectively.}
\vspace{3mm}
\label{finger_mixed_scenarios}
\begin{tabular}{c c c| c c| c c}
\multicolumn{1}{c|}{\textbf{Training-Testing}}  & \multicolumn{2}{c|}{\textbf{Low Complexity}}                                                                               & \multicolumn{2}{c|}{\textbf{Medium Complexity}}                                                                             & \multicolumn{2}{c}{\textbf{High Complexity}}                                                                                \\ \multicolumn{1}{c|}{} 
                                    & \textbf{Baseline}                                           & \textbf{Proposed}                                            & \textbf{Baseline}                                           & \textbf{Proposed}                                           & \textbf{Baseline}                                           & \textbf{Proposed}                                           \\ \hline
\multicolumn{1}{c|}{Pen-Pen} & \begin{tabular}[c]{@{}c@{}}11.1\\ 0.1\end{tabular} & \begin{tabular}[c]{@{}c@{}}8.3\\ 0.1\end{tabular}  & \begin{tabular}[c]{@{}c@{}}8.3\\ 0.1\end{tabular}  & \begin{tabular}[c]{@{}c@{}}10.2\\ 0.1\end{tabular} & \begin{tabular}[c]{@{}c@{}}5.6\\ 0.1\end{tabular}  & \begin{tabular}[c]{@{}c@{}}5.6\\ 0.1\end{tabular} \\ \hline
\multicolumn{1}{c|}{Finger-Finger} & \begin{tabular}[c]{@{}c@{}}16.7\\ 0.1\end{tabular} & \begin{tabular}[c]{@{}c@{}}11.1\\ 0.1\end{tabular} & \begin{tabular}[c]{@{}c@{}}19.4\\ 0.1\end{tabular} & \begin{tabular}[c]{@{}c@{}}15.7\\ 0.1\end{tabular} & \begin{tabular}[c]{@{}c@{}}11.1\\ 0.1\end{tabular} & \begin{tabular}[c]{@{}c@{}}10.2\\ 0.1\end{tabular}  \\ \hline
\multicolumn{1}{c|}{Pen-Finger} & \begin{tabular}[c]{@{}c@{}}30.6\\ 0.1\end{tabular} & \begin{tabular}[c]{@{}c@{}}27.8\\ 0.1\end{tabular} & \begin{tabular}[c]{@{}c@{}}22.2\\ 0.1\end{tabular} & \begin{tabular}[c]{@{}c@{}}16.7\\ 0.1\end{tabular} & \begin{tabular}[c]{@{}c@{}}11.1\\ 0.1\end{tabular} & \begin{tabular}[c]{@{}c@{}}11.1\\ 0.1\end{tabular} \\ \hline
\multicolumn{1}{c|}{Finger-Pen} & \begin{tabular}[c]{@{}c@{}}27.8\\ 0.1\end{tabular} & \begin{tabular}[c]{@{}c@{}}25.0\\ 0.1\end{tabular} & \begin{tabular}[c]{@{}c@{}}19.4\\ 0.1\end{tabular} & \begin{tabular}[c]{@{}c@{}}16.7\\ 0.1\end{tabular} & \begin{tabular}[c]{@{}c@{}}25.0\\ 0.1\end{tabular} & \begin{tabular}[c]{@{}c@{}}11.1\\ 0.1\end{tabular} \\ \hline
\end{tabular}
\end{table*}

\begin{figure}[t]
     \centering
     \subfigure[Pen]{\label{fig:pen_signature}
     \includegraphics[width=0.7\linewidth]{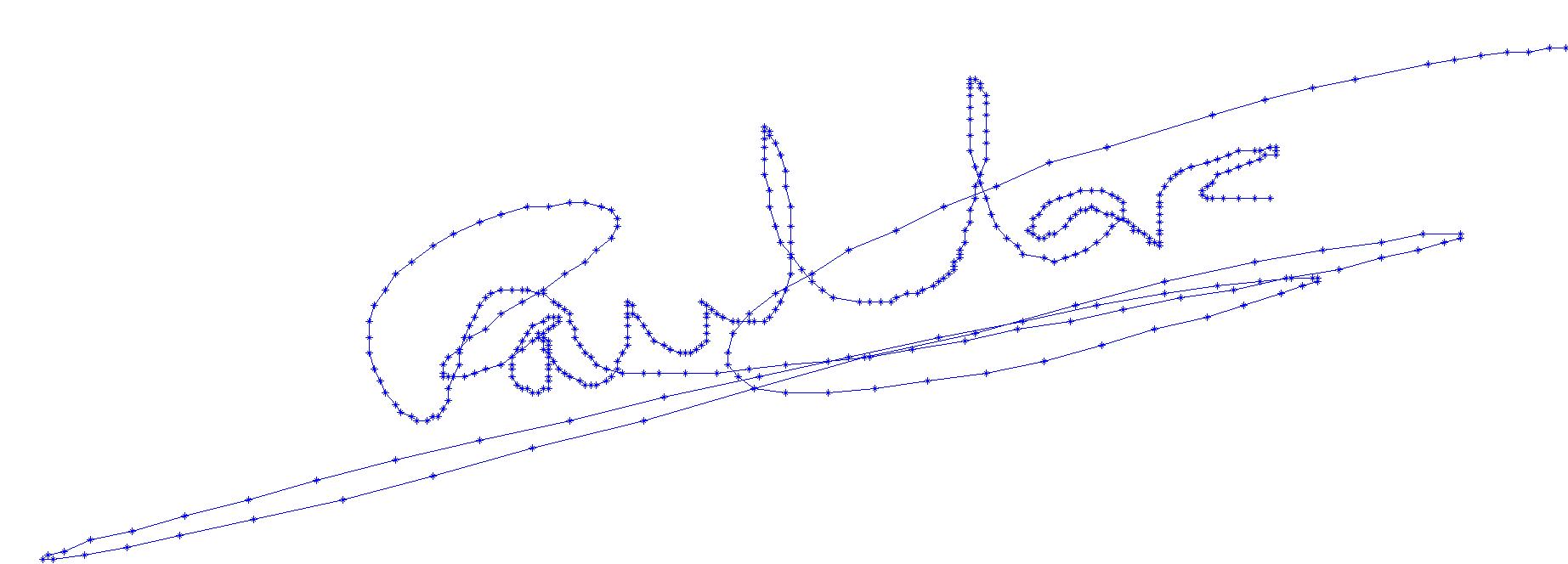}}
      \subfigure[Finger]{\label{fig:finger_signature}
     \includegraphics[width=0.7\linewidth]{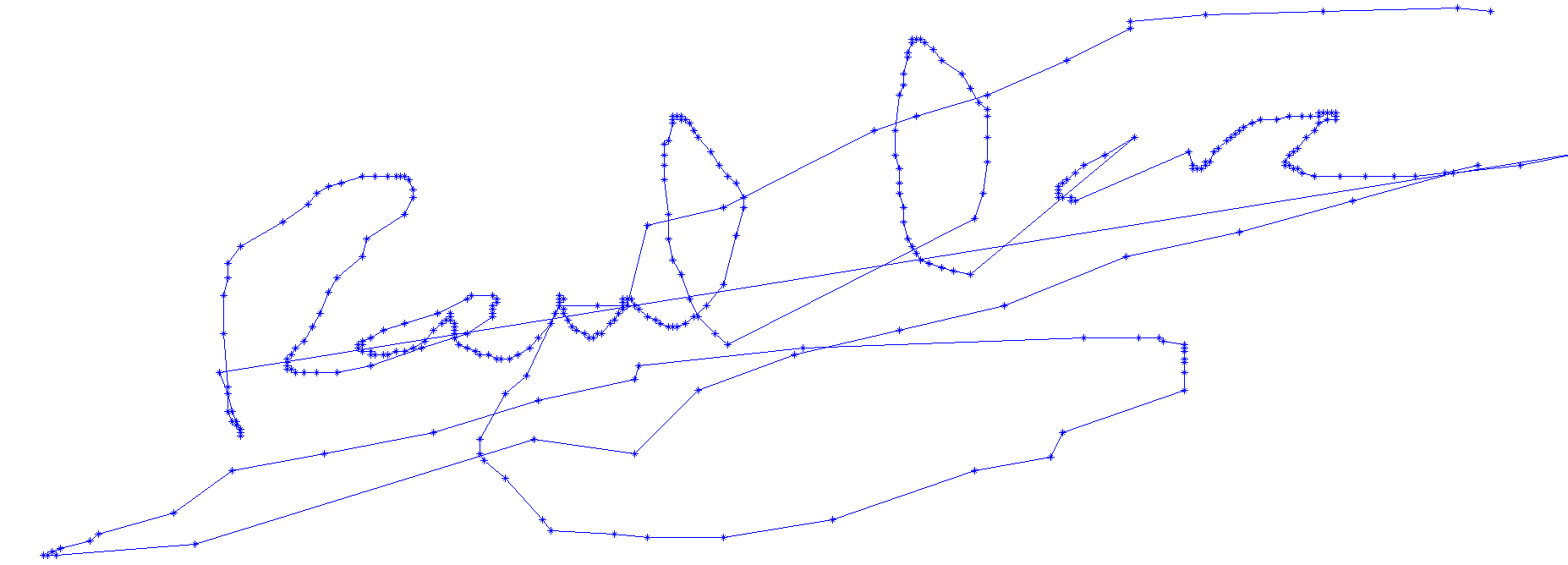}}
      \caption{Pen and finger signatures from the e-BioSign database.}
     \label{fig:pen_finger_signatures}
\end{figure}

Despite the considerable improvements achieved in the finger scenario with our proposed approach, there is still a high difference in the system performance between pen and finger scenarios (Pen-Pen vs. Finger-Finger). Concretely, an absolute worsening of 4.3\% EER is observed compared with the pen scenario. 

In order to find out the reason for such difference in the system performance, an exhaustive analysis of the finger scenario is carried out. In general, users who perform their signatures using closed letters (i.e., a, e, o, l, p, q, etc.) tend to perform much larger writing executions in comparison with other letters due to the lower precision they are able to achieve using the finger. Besides, users whose signatures are composed of a long name and surname (or two surnames) tend to simplify some parts of their signatures while signing with the finger. Regarding the sampling frequency of the acquisition process, it is important to highlight the differences between the pen and finger scenarios. For the pen scenario, all samples of the signature are uniformly distributed across the whole signing process. However, for the finger input scenario, most samples are distributed in small parts of the signature instead of the whole signature as it happens in the pen scenario. This non-desirable effect is produced due to the lack of precision achieved by the finger and also by the friction produced between the screen and the finger. Therefore, it might not be related to the specific device considered in the experimental work, but to the finger input scenario instead. 

Fig. \ref{fig:pen_finger_signatures} depicts some of the effects commented before between pen and finger scenarios. Despite this effect, and although the number of samples are very similar in both scenarios, an additional interpolation step based on splines is required to reduce the differences between the pen and finger scenarios. 

Finally, it is important to remark the challenging impostor scenario considered in this study as forgers had access to the dynamic realization of the signatures to forge. A recommendation for the usage of signature recognition on mobile devices would be for the users to protect themselves from other people that could be watching while signing, as this is more feasible to do in a mobile scenario compared with an office scenario. This way impostors might have access to the global shape of a signature but not to the dynamic information. To summarise, the higher intra-user variability together with the challenging skilled forgery scenario considered for finger input result in a degradation of the system performance compared with the pen scenario, especially for users with medium and high complexity levels, as depicted in Table \ref{finger_mixed_scenarios}.

\begin{figure*}[t]
     \centering
     \subfigure[Finger-Finger]{\label{fig:BiosecurID_pen}
     \includegraphics[width=0.32\linewidth]{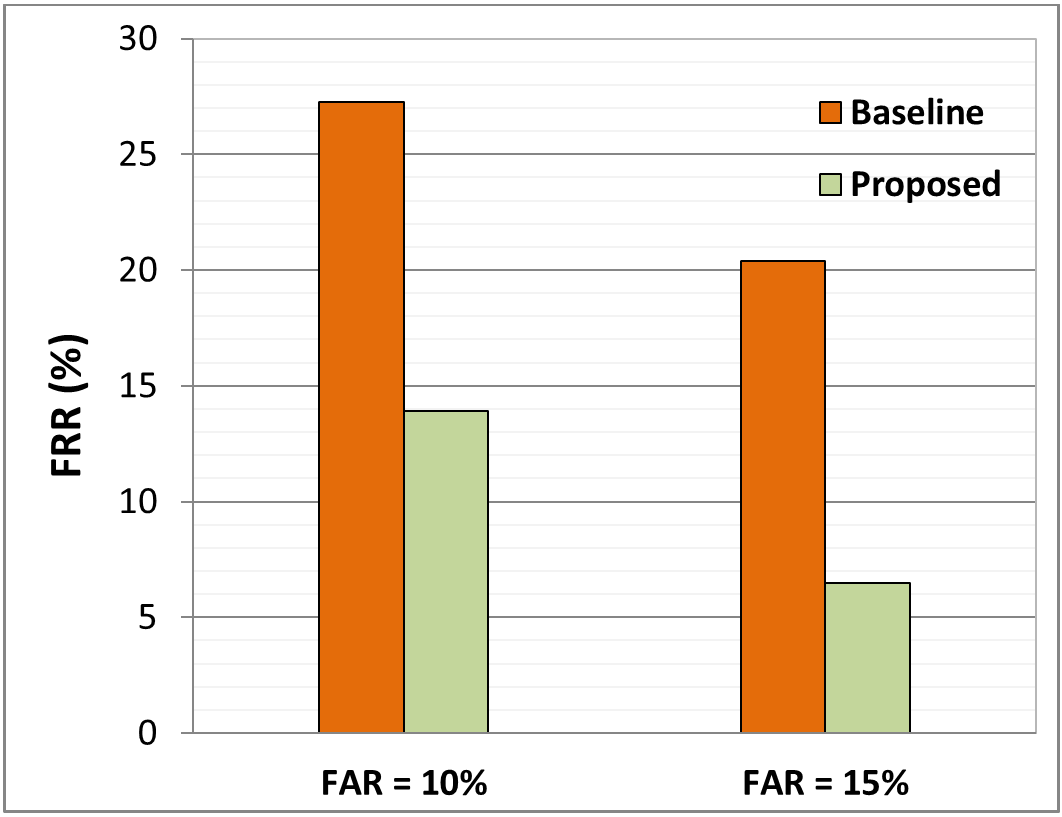}}
      \subfigure[Pen-Finger]{\label{fig:eBioSign_pen}
     \includegraphics[width=0.32\linewidth]{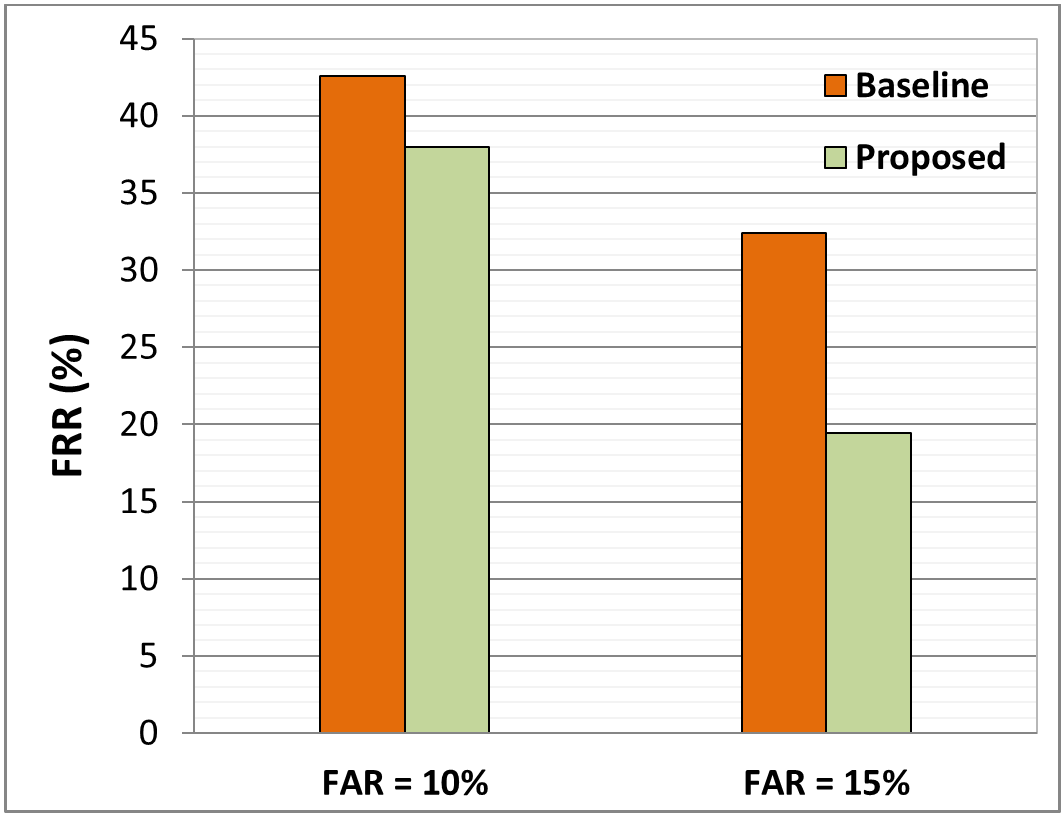}}
      \subfigure[Finger-Pen]{\label{fig:eBioSign_pen}
     \includegraphics[width=0.32\linewidth]{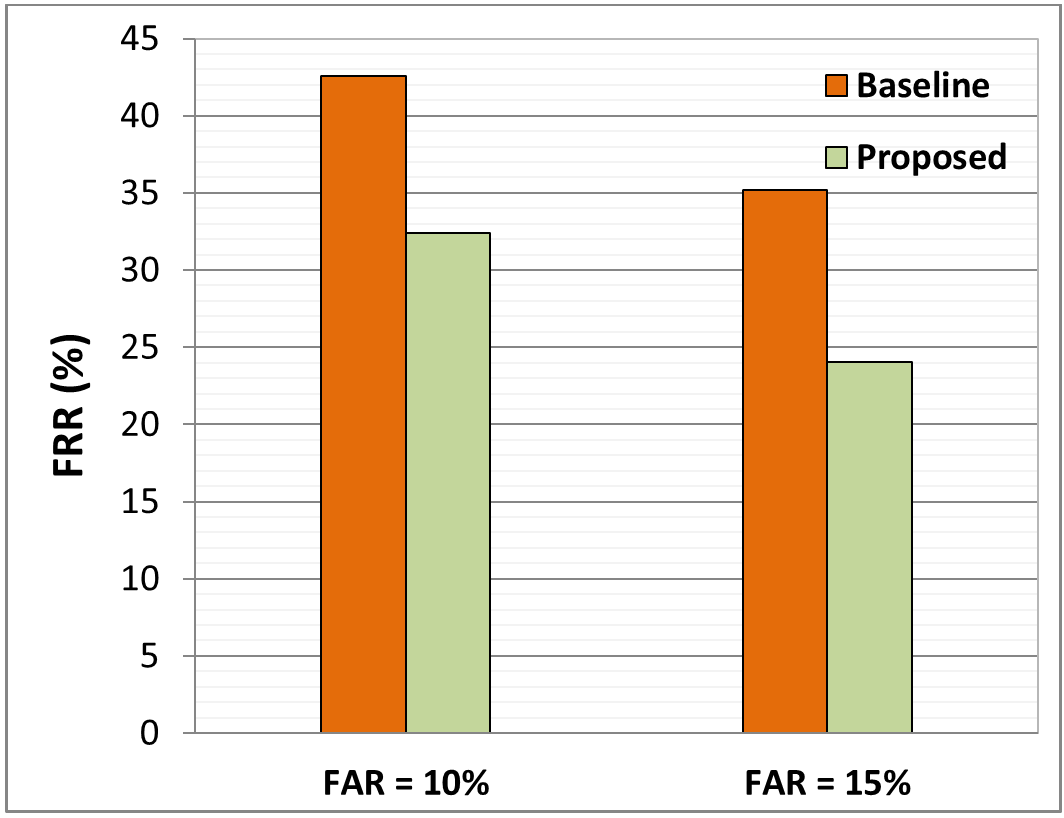}}
     \caption{\textbf{Finger and mixed input scenarios:} system performance results in terms of FRR at different values of FAR using the evaluation dataset of e-BioSign.}
     \label{fig:finger_BiosecurID_eBioSign}
\end{figure*}

\subsubsection{Mixed Input Scenarios}\label{sec:mixed_input_scenarioos}

We now study mixed input scenarios (i.e., Pen-Finger and Finger-Pen), where signatures acquired using pen and finger inputs are independently considered for training and testing the system. 

Analysing the skilled forgery results obtained in Table \ref{finger_mixed_scenarios}, our Proposed System achieves an average absolute improvement of 2.8\% and 6.5\% EER compared with the Baseline System for the Pen-Finger and Finger-Pen scenarios, respectively. It is important to remark the significant worsening of the system performance for those users with a low-complexity level with results around 25.0\% EER. Besides, these error rates are also much higher compared with the case of using the same writing input for training and testing. However, for users with medium and high complexity levels, the system performance on mixed input scenarios are very close to the Finger-Finger scenario with 16.7\% and 11.1\% EERs for medium- and high-complexity levels, respectively.

As a result, two key conclusions can be extracted from our analysis. The first one is that mixed input scenarios are feasible in practical applications for users with medium and high complexity levels. Users categorised into a low-complexity level should perform a more robust signature in order to be able to use mixed input scenarios. The second one is that the degradation of the system performance on mixed input scenarios seems to disappear for those users with medium- and high-complexity levels after considering our proposed approach, obtaining similar results to the Finger-Finger scenario. 

Finally, Fig. \ref{fig:finger_BiosecurID_eBioSign} shows the final system performance results in terms of FRR at different values of FAR for the finger and mixed input scenarios considering all complexity levels together. For the Finger-Finger scenario, our Proposed System achieves an average absolute improvement of 13.6\% FRR compared with the Baseline System, with a final value of 13.9\% FRR for a FAR = 10.0\%. For the mixed input scenarios, our Proposed System achieves an average absolute improvement of 8.8\% and 10.7\% FRR for the Pen-Finger and Finger-Pen scenarios, respectively. It is important to note the considerable improvements achieved by our proposed approach on both finger and mixed input scenarios, remarking how important the signature complexity is on these challenging scenarios. Final values of 19.4\% and 24.0\% FRR are achieved for the Pen-Finger and Finger-Pen scenarios for a value of FAR = 15.0\%. Therefore, the deployment of these scenarios on real applications seems to be more feasible with rates below 20.0\% of FRR and FAR. For a further improvement of security, a possible recommendation could be to ask clients to perform their signatures using both pen and finger writing tools during the enrolment stage in order to obtain better results, or at least for those users with low complexity.

\section{Conclusions}\label{sec:conclusions}
This paper proposes the first methodology focused on the development of an on-line signature verification system adapted to the signature complexity level. Two main modules are proposed: \textit{i)} a signature complexity detector, and \textit{ii)} a complexity-based time functions extractor.

Our proposed approach has been tested in pen, finger and mixed input scenarios considering two different on-line signature databases, BiosecurID (only for the pen scenario) and e-BioSign (for both pen and finger scenarios) with a total of 400 and 65 users, respectively. Additionally, a review of the most relevant and recent studies of signature complexity, and pen- and touch-based scenarios has been carried out in order to make our proposed approach easily comparable with previous studies.

Analysing the results obtained for the pen scenario, our Proposed System has achieved for the BiosecurID database an average absolute improvement of 2.5\% EER for skilled forgeries compared with the Baseline System. This improvement has been even higher for the finger scenario, achieving an average absolute improvement of 3.4\% EER for the e-BioSign database. Additionally, our Proposed System has achieved for the finger scenario an absolute improvement of 5.6\% EER for the most challenging users (i.e., users with a low complexity-level as they are easier to forge). All these improvements prove the success of our proposed approach based on the signature complexity.

For future work, new approaches based on Recurrent Neural Networks~\cite{2018_IEEEAccess_RNN_Tolosana} will be studied in order to \textit{i)} develop more accurate signature complexity detectors, and \textit{ii)} reduce the system performance degradation on these thriving but challenging scenarios. Also, unsupervised techniques will be studied in order to exploit large-scale signature datasets not manageable for human labelling.

\section*{Acknowledgments}
This work has been supported by projects: BIBECA (RTI2018-101248-B-I00 MINECO/FEDER), Bio-Guard (Ayudas Fundaci\'on BBVA a Equipos de Investigaci\'on Cient\'ifica 2017) and by UAM-CecaBank.

{
\bibliographystyle{IEEEtran}
\bibliography{egbib2}
}

\end{document}